==================
\documentstyle[preprint,revtex]{aps}

\newcommand{\bphi}{{\bar\phi}}

\newcommand{\KMM}{Kazakov--Migdal~model~}

\newcommand{\bq}{\begin{eqnarray}}
\newcommand{\eq}{\end{eqnarray}}
\newcommand{\be}{\begin{equation}}
\newcommand{\ee}{\end{equation}}
\newcommand{\ba}{\begin{eqnarray}}

\newcommand{\ea}{\end{eqnarray}}
\newcommand{\baa}{\begin{eqnarray*}}
\newcommand{\eaa}{\end{eqnarray*}}
\newcommand{\barr}{\begin{array}}
\newcommand{\earr}{\end{array}}
\newcommand{\bb}{\begin{thebibliography}}
\newcommand{\eb}{\end{thebibliography}}

\newcommand{\lab}[1]{\label{#1}}

\newcommand{\Tr}{\mbox{Tr\,}}

\newcommand{\de}{\delta}
\newcommand{\ad}{\dagger}

\def\vsmall{\@setsize\vsmall{10pt}\vspt\@vspt}

\begin{document}
\draft

\preprint{ PUPT 1358, UBCTP92-032 \\December 2 1992}
%\nopagenumbers

\begin{title}
{\bf  Induced QCD Without Local Confinement}
\end{title}

%%%%
\vskip .5in
%%%%

\author{\bf M.I.Dobroliubov$^{1,a}$,
I.I.Kogan$^{2,b}$,
 G.W.Semenoff$^1$ and N.Weiss$^1$}

%%%%
\vskip 15pt
%%%%

\begin{instit}
1. Department of Physics, University of British Columbia,\\ Vancouver,
British Columba, Canada V6T 1Z1\\
2. Department of Physics, Princeton University,\\ Princeton, New
Jersey 08544 U.S.A.
\end{instit}
%%%%

\begin{abstract}
\vskip .2 truein
We examine some properties of the filled Wilson loop observables in
the Kazakov-Migdal model of induced QCD. We show that they have a
natural interpretation in a modification of the original model in
which the $Z_N$ gauge symmetry is broken explicitly by a Wilson
kinetic term for the gauge fields.  We argue that there are two large
N limits of this theory, one leads to ordinary Wilson lattice gauge
theory coupled to a dynamical scalar field and the other leads to a
version of the Kazakov-Migdal model in which the large N solution
found by Migdal can still be used.  We discuss the properties of the
string theory which emerges.

\vskip 1.6 truein

\hrule
\vskip .2in
\noindent (a) Permanent Address:
 Institute for Nuclear Research, Academy of Sciences of Russia, 60TH
October Anniversary Prospekt 7A, Moscow 117312, Russia.

\noindent (b) Permanent Address:
 Institute for Theoretical and Experimental Physics, Bolshaya
Cheremushkinskaya, 117259 Moscow, Russia.

\end{abstract}
\newpage

The dynamics of QCD is known to simplify somewhat in the limit where
the number of colors, $N$, is large \cite{lgn}.  Only planar graphs
contribute to scattering amplitudes and the resulting perturbation
series exhibits some of the qualitative features of the strong
interactions.  However so far, no explicit solution is available in
the large $N$ limit and it has thus led to very few quantitative
results.

Recently, Kazakov and Migdal \cite{kmm} have proposed a novel approach
which has the hope of providing an exact solution of the large $N$
limit of QCD.  They consider induced QCD which is obtained by
integrating over the scalar fields in the lattice gauge theory with
the partition function
\equation
	Z_{\rm KM}=\int d\phi [dU]\exp\left(-N\sum_x{\rm
	Tr}V[\phi(x)]+N\sum_{<x,y>}{\rm
	Tr}\phi(x)U(xy)\phi(y)U^{\dagger}(xy)\right)
\label{km}
\endequation
where $\phi(x)$ are N$\times$N Hermitean matrices which reside on
lattice sites x, $U(xy)$ are unitary N$\times$N matrices which reside
on links $<xy>$ between neighboring sites x and y, $d\phi$ is the
Euclidean integration measure for Hermitean matrices, $[dU]$ is the
invariant Haar measure for integration over the unitary group U(N) and
$V[\phi]$ is a potential for the scalars.  This model is invariant
under the gauge transformations $\phi(x)\rightarrow
\omega(x)\phi(x)\omega^{\dagger}(x)$, $ U(xy)\rightarrow
\omega(x)U(xy)\omega^{\dagger}(y) $ where $\omega(x)$ is an element of
U(N).  By restricting the trace of $\phi$ to zero and the determinant
of $U$ to one in (\ref{km}) we could also consider a model with SU(N)
gauge symmetry.

This model is soluble in the large $N$ limit.  By explicitly
integrating over the unitary matrices in (\ref{km}) (see Appendix A)
one obtains a field theory for the eigenvalues of $\phi$
\equation
	Z_{\rm KM}\propto\int\prod_{x,i} d\phi_i(x)\Delta^2[\phi(x)]
e^{-N\sum_x V[\phi_i(x)]}\prod_{<xy>}
{\det_{ij}e^{N\phi_i(x)\phi_j(y)}\over\Delta[\phi(x)]\Delta[\phi(y)]}
\label{phii}
\endequation
where
$\Delta[\phi]=\det_{ij}(\phi_i)^{j-1}=\prod_{i<j}(\phi_i-\phi_j)$ is
the Vandermonde determinant for $\phi$.  The eigenvalues $\phi_i$
behave like a master field since the large $N$ limit in (\ref{phii})
is the classical limit and the integral can be performed by saddle
point approximation.  Migdal \cite{mig} has given an expression for
the saddle point in terms of the density of eigenvalues.  Corrections
to the classical behavior and the spectrum of elementary excitations
can also be computed \cite{mig2}. This model has been considered
further in
\cite{ksw} - \cite{boulatov2}.

If the model (\ref{km}) has a second order phase transition and if
the fluctuations in the vicinity of the critical point are
non-Gaussian, one might expect that the critical behavior should be
represented by QCD, the only known nontrivial four dimensional field
theory with non-Abelian gauge symmetry.  However it was pointed out in
\cite{ksw} that, like all adjoint lattice models, the Kazakov-Migdal
model has an extra gauge symmetry which is not a symmetry of continuum
QCD.  The action in (\ref{km}) is invariant under redefining any of
the gauge matrices by an element of the center of the gauge group,
$U(xy)\rightarrow z(xy)U(xy)$, $\phi(x)\rightarrow\phi(x) $, where
$z(xy)\in U(1)$ if the gauge group is $U(N)$ and $z(xy)\in Z_N$ if the
gauge group is SU(N). (We shall call the symmetry a $Z_N$ gauge
symmetry in either case.  This symmetry is the maximal subgroup of the
transformations discussed by Gross\cite{Gr92} and
Boulatov\cite{boulatov} which can be implemented with field
independent elements $z(xy)$.) Because of this symmetry the
conventional Wilson loop observables of lattice gauge theory have
vanishing average unless they have either equal numbers of $U$ and
$U^{\dagger}$ operators on each link or else, in the case of SU(N),
unless they have an integer multiple of $N$ $U$'s or $N$
$U^{\dagger}$'s.

In conventional QCD, the expectation value of the Wilson loop operator
gives the free energy for a process which creates a heavy
quark-antiquark pair, separates them for some time and lets them
annihilate.  From the asymptotics for large loops, one extracts the
interaction potential for the quarks.  If the expectation value of the
Wilson loop behaves asymptotically like $e^{-\alpha A}$ where A is the
area of a minimal surface whose boundary is the loop, the
quark-antiquark potential grows linearly with separation at large
distances and quarks are confined. The parameter $\alpha$ is the
string tension. On the other hand if the expectation value of the
Wilson loop goes like the exponential its perimeter then the potential
is not confining.  In the Kazakov-Migdal model, due to the $Z_N$
symmetry, the expectation value of the Wilson Loop is identically zero
for all loops with non-zero area.  We can interpret this as giving an
area law with $\alpha=\infty$, and no propagation of colored objects
is allowed at all. (An exception is the baryon ($U^N$) loops in the
case of SU(N) where the correct statement is that $N$-ality cannot
propagate.)  It is for this reason that the original Kazakov-Migdal
model has difficulty describing pure gluo-dynamics.

There are currently several points of view on how to cure this
difficulty.  In \cite{ksw} it was suggested that if there is a phase
transition so that the $Z_N$ symmetry is represented in a Higgs phase,
the resulting large distance theory would resemble conventional QCD.
This approach has been pursued in \cite{KhM92},\cite{M92},\cite{KM92}.
An alternative, which was advocated in \cite{ksw,kmsw1}, is to use
unconventional observables such as filled Wilson loops which reduce to
the usual Wilson loop in the naive continuum limit but which are
invariant under $Z_N$.  The third possibility is to break the $Z_N$
symmetry explicitly.  This was suggested by Migdal
\cite{mixed} in his  mixed model in which he breaks the $Z_N$ symmetry
by introducing into the model heavy quarks in the fundamental
representation of the gauge group.  In this Letter we shall examine a
different explicit $Z_N$ symmetry breaking and the use of filled
Wilson loop observables.  We shall show that the filled Wilson loops
arise naturally from ordinary Wilson loops in a modified version of
the Kazakov-Migdal model which has additional explicit symmetry
breaking terms.  We shall argue that one version of this modified
model should be solvable in the large $N$ limit. It thus provides a
link between the approach which uses unconventional observables to
solve the $Z_N$ problem and the approach in which the symmetry is
explicitly broken.

We begin with a brief description of the filled Wilson loop operators
which were introduced in \cite{ksw} and discussed in detail in
\cite{kmsw1}.  These are a special class of correlation functions
which survive the $Z_N$ symmetry. They are defined by considering an
oriented closed curve $\Gamma$ made of links of the lattice.  The
ordinary Wilson loop operator on $\Gamma$ is given by
\equation
	W[\Gamma]={\rm
	Tr}\left\{\prod_{<xy>\in\Gamma}U(xy)\right\}.
\endequation
For any surface $S$ which is made of plaquettes such that the boundary
of $S$ is the curve $\Gamma$ we define
\equation
	W_F[\Gamma,S]=
	W[\Gamma]\prod_{\Box\in S}W^{\dagger}[\Box]
\endequation
where $\Box$ denotes an elementary plaquette in the surface $S$. The
filled Wilson loop for $\Gamma$ is now defined as
\equation
	W_F[\Gamma]=\sum_S\mu(S)W_F[\Gamma,S]
\endequation
where the sum is over all
surfaces $S$ whose boundary
is the loop $\Gamma$ with some (yet to be specified) weight function
$\mu(S)$.  Notice that for each plaquette $\Box\in S$ we have inserted
the negatively oriented Wilson loop $W^{\dagger}[\Box]$.
Thus for arbitrary weight functional $\mu(S)$ the filled Wilson loop
operator is invariant under the local $Z_N$ gauge symmetry
since it  has equal numbers of U
and U$^{\dagger}$ operators on each link.
Although we have assumed that the loop is filled with
{\it elementary} plaquettes this can be easily generalized to other
fillings (the other extreme case being the adjoint loop
$W[\Gamma]W^{\dagger}[\Gamma]$). We can also define the
`filled correlator'' of more than one
loop by summing over all surfaces whose boundary is given
by those loops.

We now review some of the properties of the filled Wilson loop.
In ref. \cite{kmsw1} it was shown that computing the expectation value
of $W_F[\Gamma]$ is equivalent to computing the partition function of
a certain statistical model on a random two--dimensional
lattice.  When computing $Z_N$ gauge invariant correlation functions
of $U$--matrices in the master field approximation the $\phi$--integral
is evaluated by substituting the master field $\bar\phi={\rm
diag}(\bar\phi_1,\ldots,\bar\phi_N)$ for the eigenvalues of $\phi$.

\equation
	<U_{i_1j_1}\dots U^{\dagger}_{k_1l_1}\dots>~=
	{\int
	d\phi[dU]e^{-{\rm Tr}(\sum V[\phi]-\sum \phi U\phi U^{\dagger})}
	U_{i_1j_1}\dots U^{\dagger}_{k_1l_1}\dots\over\int
	d\phi[dU]e^{-{\rm Tr}(\sum V[\phi]-\sum \phi U\phi U^{\dagger})}}
\label{correl}
\endequation
$$
	\approx{\int
	d\bar\phi[dU]e^{{\rm Tr}(\sum \bar\phi
	U\bar\phi U^{\dagger})}
	U_{i_1j_1}\dots
	U^{\dagger}_{k_1l_1}\dots\over\int
	d\bar\phi[dU]e^{{\rm Tr}(\sum
	\bar\phi U\bar\phi U^{\dagger})}}
$$
If we consider for the moment
surfaces which are not self-intersecting so that the filled Wilson
loop correlator has at most one $UU^{\dagger}$ pair on any link we
need to consider only the two field correlator
$<U_{ij}U_{kl}^{\dagger} >$. Gauge invariance implies that\cite{kmsw1}
\equation
	<U_{ij}U_{kl}^{\dagger} >=C_{ij}\delta_{il}\delta_{jk}~~~{\rm with}
	~~~C_{ij}={\int [dU]e^{N\sum{\rm Tr}\left(
	\bar\phi U\bar\phi U^{\dagger}\right)}
	\vert U_{ij}\vert^2\over\int [dU]e^{N\sum{\rm Tr}\left(
	\bar\phi U\bar\phi
	U^{\dagger}\right)}}
\endequation
Thus, in the master field approximation, the
expectation value of the filled Wilson loop is given by
\equation
	<W_F[\Gamma]>=\sum_S\mu(S) \left[\left(\prod_{\buildrel{
	{\rm sites}}\over{x\in S}}
	\sum_{i(x)=1}^N\right)\prod_{\buildrel
	{\rm links}\over{<xy>\in S}}C_{i(x)~j(x)}\right]
\label{fwl}
\endequation
This is a generalized Potts
model on a random surface in which $N$--component spins reside at each site
and the Boltzmann weights $C_{ij}$ for the bonds are correctly
normalized to be
conditional probabilities; $\sum_i C_{ij}$=$1$, $\sum_j C_{ij}$=$1$.

Techniques for evaluating $C_{ij}$ for general $N$ and for arbitrary
$\bar\phi$ are presented in \cite{shat} and \cite{moroz}. An explicit formula
for $SU(2)$ is given in \cite{kmsw1}. Although the general formula for
$C_{ij}$ in $SU(N)$ is quite difficult to deal with,
it is still possible to estimate the surface
dependence of the statistical model partition function in (\ref{fwl})
when $\bar\phi$ is homogeneous by considering two different limits.
(The details of the computation of $C_{ij}$ in these limits is
described in the Appendix.)  First is the limit in which
$\bar\phi$ is small. We call this the ``High Temperature'' limit since
in this limit
\equation
	 C_{ij}^{\mbox{\tiny\bf HT}}={1\over N}+\dots
\endequation
is independent of $i$ and $j$. It thus represents the Bolzman weights
for a highly disordered system. We also consider the limit in
which $\bar\phi$ is large. We call this the ``Low Temperature'' limit
since in this limit
\equation
	C_{ij}^{\mbox{\tiny\bf LT}}=\delta_{ij}+\dots
\endequation
and the value of the spin at each site is equal. In this case the
$C_{ij}$ represent the Bolzman weight for a perfectly ordered system.
These two cases lead to profoundly different
behavior for the filled Wilson loop.  We shall assume that, by choosing
the potential for the scalar field in (\ref{phii}) appropriately, either
of these limits could be obtained (the eigenvalue repulsion due to the
Vandermonde determinants in (\ref{phii}) and the possibility of adding
repulsive central potentials makes the low temperature limit more natural).

We begin by estimating the value of the filled Wilson loop for a fixed
surface $S$.  In the ``high temperature'' case the statistical model
is disordered.  The sums over configurations at the various sites are
independent and they contribute an overall factor $N^{V}$ (where $V$
is the number of vertices on the surface) to the expectation value of
the filled Wilson loop.  Furthermore each link contributes a factor
$C_{ij}$$=$$1/N$ so that the links contribute a total factor of
$N^{-L}$ where $L$ is the total number of links. It follows that the
expectation value of the filled Wilson loop goes like
\equation
	<W_F[\Gamma,S]>^{\bf HT}\sim N^{V-L} = N^{2-2g(S)} N^{-A(S)}
\endequation
 where $A(S)$ is the area and $g(S)$ is the genus of the surface $S$
(i.e. the number of plaquettes comprising $S$) and we have used
Euler's theorem, $\chi \equiv 2-2g = V -L + A$.  We thus get the
renormalization of the string tension $\delta\alpha_{\bf HT} =\log N$.
Notice also that higher genus surfaces are suppressed and that the
loop (genus) expansion parameter is $1/N^{2}$.  This is precisely what
is obtained in the conventional strong coupling expansion of Wilson's
lattice gauge theory which is known to describe a string theory with
extra degrees of freedom associated with self-intersections of the
string
\cite{kazak}.

In the ``low temperature'' case, the statistical system is ordered.
The spins on all  the sites are frozen at a uniform value.
In this case  the partition
function is proportional to the degeneracy of the ground state,
\equation
	<W_F[\Gamma,S]>^{\mbox{\tiny\bf LT}}=N
\endequation
Note that in this case the statistical model
gives no contribution to the string tension  ($\delta\alpha_{\mbox{\tiny\bf
LT}}
\approx 0$) and there is no suppression of higher genus surfaces.

In order to proceed to the evaluation of the filled Wilson loop we
need to choose a weight function $\mu(S)$  in order to perform the sum
over surfaces. The most reasonable criterion for choosing such
a weight function is our desire
to get a finite physical string tension in the continuum
limit. In order to accomplish this goal we must choose a weight
function $\mu(S)$ which depend
on the area of the surface differently in the low and in the high
temperature cases.  It is
known that the number of closed surfaces with a given area grows
exponentially as
\equation
	n(A)\sim A^{\kappa(g)}e^{\mu_0 A}
\endequation
where $\kappa(g)$ is a universal constant which depends only on the
genus of the surface and $\mu_0$ is a non-universal, regulator
dependent constant\cite{zam} which will lead to a
renormalization of the string tension.  In our case, although the
surfaces are open, the above formula should still be valid for
surfaces whose area is much larger than the area of the minimal
surface bounded by $\Gamma$.  If the continuum limit of our theory is
realized in the ``high temperature'' phase we should use the weight
function $\mu_{\mbox{\tiny\bf HT}}(S)\sim N^{A(S)}e^{-\mu_0A(S)}$. This leads
to a
vanishing string tension in the lattice theory which is a necessary
condition for having a finite string tension in the continuum limit.
To accomplish the same goal in the ``low temperature'' phase we should
use $\mu_{\mbox{\tiny\bf LT}}(S)\sim e^{-\mu_0A(S)}$. Although these choices of
$\mu(S)$ give the desired result, it is rather unnatural to have to
choose $\mu(S)$ in such an ad hoc fashion.

Fortunately there is a
very natural way to obtain the sum over surfaces in (\ref{fwl}).
Consider the following expectation value
\equation
	<W_F[\Gamma]>={ <W
	[\Gamma]e^{\lambda\sum_\Box\left(W[\Box]+W^{\dagger}[\Box]\right)}>
	\over <e^{\lambda\sum_{\Box}\left(W(\Box)+W^{\dagger}[\Box]\right)}>}
	\label{nfwl}
\endequation
where $W[\Gamma]$ is the conventional Wilson loop. Remember that the
average is weighted by the Kazakov--Migdal action as in
(\ref{correl}): In the master field limit it is computed by
integrating only over $U$--matrices with $\phi=\bar\phi$ and with the
Kazakov--Migdal action. Note that the exponent in (\ref{nfwl}) is
simply the conventional Wilson kinetic term for the gauge fields in
lattice gauge theory. If we expand the right hand side of (\ref{nfwl})
in $\lambda$ the non--vanishing terms are all of those surfaces which
fill the Wilson loop.  The result is thus a filled Wilson loop with a
surface weight $\mu(S)=\lambda^{A(S)}$.  It is clear that we could
obtain exactly the same expression (in the master field approximation)
by evaluating the expectation value of the {\it ordinary} Wilson loop
operator in a modified version of the Kazakov--Migdal model in which a
conventional Wilson term
($\lambda\sum_{\Box}\left(W(\Box)+W^{\dagger}[\Box]\right)$) is added
to the action.  This term breaks the $Z_N$ gauge symmetry explicitly
and allows Wilson loop operators with non--zero area to have non--zero
expectation values. We would expect that it is necessary to keep
$\lambda$ small if one is to maintain the successes of the
Kazakov--Migdal model.  We shall now argue that in the ``low
temperature'' limit this picture is self--consistent in the sense that
the physical string tension is finite when $\lambda$ is small and
consequently the saddle point solution of the original model is
unchanged.  We shall also see that this is not the case in the ``high
temperature'' phase.

Let us begin by determining how $\lambda$ should behave in the
continuum limit if we are to have a finite physical string tension.
As discussed above a necessary condition for having a finite physical
string tension is that the string tension {\it in lattice units}
should vanish. It is thus necessary for the bare string tension
$-\ln\lambda$ to be chosen so as to precisely cancel the
renormalization of the string tension due to both the statistical
model to the sum over surfaces.  It is straightforward to check that
in the ``high temperature'' phase we must choose $\lambda_{\bf
HT}=Ne^{-\mu_0}$, whereas in the ``low temperature'' phase we must
choose$\lambda_{\mbox{\tiny\bf LT}}=e^{-\mu_0}$.  Notice that this $\lambda$ is
proportional to $N$ in the ``high temperature'' phase and thus cannot
be assumed small in large $N$.

In the large $N$ limit of conventional lattice gauge theory the
coefficient of the Wilson term must be proportional to $N$ if one is
to obtain a consistent large $N$ expansion.  In our case we see that
this is true for the ``high temperature'' phase in which case the
Wilson term is of the same order as the Kazakov--Migdal term and it
thus plays an important role in the infinite $N$ limit.  One can say
that in this phase we have ordinary QCD.  Unfortunately it is
impossible to preserve the master field solution of the Kazakov-Migdal
model in this limit since the Wilson term, being of order $N$, would
modify the large N solution, ruining the self-consistency of the
mean-field approximation as described here.

The situation is much more appealing in the ``low temperature'' phase.
In this case the required coefficient of the Wilson term is of order
one. It is subdominant and therefore negligible in the large $N$
limit.  Thus, Migdal's solution \cite{mig} of the Kazakov-Migdal model
in the large $N$ limit should still apply to our proposed modification
of the action.  In fact the only reason that the Wilson term is
important at all in the large $N$ limit of the ``low temperature''
phase is related to the collective phenomenon which orders the
statistical system on the surfaces.  It effectively makes the
statistical model's contribution to the string tension much smaller
than would be expected from naive counting of powers of 1/N and a
truly infinitesimal breaking of the $Z_N$ gauge symmetry
($\lambda_{\mbox{\tiny\bf LT}}/N\rightarrow0$ as $N
\rightarrow\infty$) is sufficient to make the averages of Wilson loop operators
non-vanishing. The self-consistency of this picture can also be
demonstrated by computing the contribution of the Wilson term to the
free energy.  This can be computed in a small $\lambda$ expansion. For
a cubic lattice the result is:
\equation
Z=<{e}^{\lambda_{\mbox{\tiny\bf LT}}\left(\sum_\Box{\rm
Tr}(W[\Box]+W^{\dagger}[\Box])\right)}> =~Z_{KM}~{\rm exp}{
\left(NV\frac{D(D-1)}{2}\left(\lambda_{\mbox{\tiny\bf
LT}}^2+2\lambda_{\mbox{\tiny\bf
LT}}^6+\ldots\right)\right)}
\label{freeenergy}
\endequation
is of order $N$ (where $V$ is the volume, $D$ is the dimension). This
should be compared with the free energy in the pure Kazakov-Migdal
model which is proportional to $N^2$.  Here, the first term in the
free energy is the contribution of the doubled elementary plaquette
and the second term is due to the two orientations of the elementary
cube.  It is interesting that, to order 6, there is no energy of
interaction of doubled elementary plaquettes with each other. We
conjecture that the interaction energy of surfaces is absent to all
orders and the free energy obtains contributions from all possible
topologically distinct surfaces which can be built from elementary
plaquettes. This suggests a free string picture of the ``low
temperature'' limit of the Kazakov-Migdal model at lattice scales.

We have thus far neglected the self-intersecting surfaces in the sum
(\ref{fwl}) which are generated by the expansion of (\ref{nfwl}) in
$\lambda$.  In order to evaluate the contribution of these surfaces we
need to compute the correlator of $n$ $UU^{\dagger}$ pairs on the same
link. The computation of these correlators in full generality is quite
complicated. In the Appendix we compute them in the ``low
temperature'' (ordered) phase.  We find that
\equation
	<U_{i_1j_1}\ldots U_{i_nj_n}U^{\dagger}_{k_1l_1}\ldots
	U^{\dagger}_{k_nl_n}> =\delta_{i_1j_1}\ldots\delta_{k_nl_n}S^{i_1\dots
	i_n}_{k_1\dots k_n}
\endequation
where $S^{i_1\dots i_n}_{k_1\dots k_n}$ is the tensor which is one if
$i_1\dots i_n$ is a permutation of $k_1\dots k_n$ and is zero
otherwise.  It is now evident that in this limit the $U$--matrices are
replaced by unit matrices which freeze together the spin degrees of
freedom on the various intersecting surfaces.  As a special case we
can consider a single, connected, self--intersecting surface. In this
case all the spin indices on the surface are equal and since
$S^{i_1\dots}_{k_1\dots}=1$ when all arguments are equal the partition
function of the statistical model corresponding to that surface is
simply $N$ just as it was for a non-intersecting surface.  Thus just
as the statistical model does not contribute to the string tension it
also does not contribute to the interaction energy of
self--intersecting surfaces.  This implies that in the ``low
temperature'' limit, the sum over connected surfaces which have a
common boundary behaves like a Nambu-Goto string theory with no
internal degrees of freedom.

In summary, the self-consistency of the ``low temperature'' limit
leads us to a new large N limit of the conventional lattice gauge
theory coupled to scalars: $$ 	Z=\int d\phi
[dU]\exp\left(-N\sum_x{\rm Tr}V[\phi(x)]+N\sum_{<x,y>}{\rm
Tr}\phi(x)U(xy)\phi(y)U^{\dagger}(xy)+\right.  $$
\equation
	\left.+
	\lambda\sum_{\Box}
	(W(\Box)+W^{\dagger}(\Box))\right)
	\label{nkmm}
\endequation
The conventional large $N$ limit occurs when $\lambda $ is of order
$N$ and describes scalar QCD.  The other limit occurs when
$N\rightarrow \infty$ with $\lambda$ of order one. This model is
soluble using the Kazakov--Migdal approach.

It is the latter case in which $\lambda$ remains constant that is of
special interest to us.  In this case we saw that the large $N$
expansion corresponds to a string theory with some unusual features.
The partition function and the Wilson loop expectation value can be
described as a sum over surfaces.  What is unusual is that the genus
of the surfaces is not suppressed in the large $N$ limit, as it is in
continuum QCD.  (We do of course expect the higher corrections in
$1/N$ to suppress higher genus terms.) For a continuum string theory
this sum over the genus is badly divergent.  This, together with the
presence of tachyons, suggests that the true ground state of the
string theory is some sort of condensate.  This could pose a
complication for the present version of the Kazakov-Migdal model in the
continuum limit and deserves further attention.  It is still a mystery
to us how the sum over all surfaces at the lattice scale should turn
into the sum over planar diagrams in the continuum theory of QCD.

An alternative to the model presented here is the mixed model which
was invented by Migdal \cite{mixed} to solve the problem of $Z_N$
symmetry.  It contains heavy quarks in the fundamental representation
of the gauge group. Despite the obvious differences between our model
and Migdal's mixed model they have many features in common.  As in all
cases when there are fields in the fundamental representation, the
asymptotics of the Wilson loops in the mixed model exhibit a perimeter
law.  In conventional QCD one would expect that if the quarks are
heavy enough, there is an area law for small enough loops, i.e. there
would exist a size scale which is far enough into the infrared region
that the quark potential is linear but the interaction energy is not
yet large enough that it is screened by producing quark-antiquark
pairs.  Thus, in QCD we expect that adding heavy quarks would not ruin
the area law for Wilson loops smaller then some scale.

The mixed model has just the opposite scenario, it is possible to get
an area law only when the heavy quarks are light enough.  This is a
result of the fact that, in the Kazakov-Migdal model, no Wilson loops
are allowed at all unless the $Z_N$ symmetry is explicitly broken.  In
the mixed model, the $Z_N$ charge of links in a Wilson loop must be
screened by the heavy quarks.  This can happen in two ways.  First,
the Wilson loop can just bind a heavy quark to form an adjoint loop -
giving a perimeter law for the free energy of the loop.  This is the
leading behavior if the fermion mass, $M$, is large. The free energy
would go like $1/M^P$ where $P$ is the perimeter.  The only way an
area law might arise is when the fermions are light enough that their
propagators could from a filled Wilson loop with free energy
$1/M^{2L}$ where $L\approx 2A$ is the number of links.  Then, since
the entropy for filled loops is much larger than that for adjoint
loops, these configurations would be important if $M^4<e^{\mu_0}$.
Then, the asymptotics behavior of the Wilson loop would still have a
perimeter law but there would be loops with $4A-P<\mu_0/\ln M$ where
there would be an approximate area law.

{\bf Acknowledgement} This work is supported in part by the Natural
Sciences and Engineering Research Council of Canada and by the
National Science Foundation grant \# NSF PHY90-21984.  We thank D.
Gross, Yu.  Makeenko, A. Migdal, A.  Morozov and S. Shatashvili for
discussions.

\vskip .4in
\noindent{\bf APPENDIX:~~~
{Asymptotics of Correlators}}
\vskip .2in

The key to the solvability of the \KMM is the fact that the single--link
Itzyksen--Zuber integral can be done analytically \cite{itzub,harish}
\equation
	I_{IZ}=\int [dU] e^{N\sum \phi_i\chi_j\vert U_{ij}\vert^2}=
	{\det_{(ij)}e^{\phi_i\chi_j}\over\Delta[\phi]\Delta[\chi]}
\label{itzub}
\endequation
In this Appendix we consider two limiting cases of this integral
and its correlators.  These have already been solved in
\cite{shat,moroz}.  Here we present a quick and
simple derivation in two limiting cases.
\bigskip

\noindent
{\bf 1. Low Temperature:~~~} First consider the case
where $\bphi_i$ are large.  We also assume that the eigenvalues
$\bphi_i$ are not too close to each other in the sense that
$	\sum_{i\neq j}
	{1\over{\left(\bphi_i-\bphi_j\right)^2}} << N$
(these two conditions are satisfied in the case of the semicircle
distribution of the eigenvalues).  The integral in (\ref{itzub}) is known
to be exact in the semi-classical approximation (see \cite{kmsw1} for a
discussion).    The classical equation of motion is $\bigl[ U\bphi U^{\dagger},
\bphi\bigr]=0$ which, since $\bphi$ is diagonal, is solved by any $U$ of
the form $U_0=DP$ where $D$ is a diagonal unitary matrix and $P$ is a
matrix which permutes the eigenvalues, $(P\bphi P)_{ij}=\delta_{ij}
\bphi_{P(i)}$.  Also, when N is large and $\bphi$ is not too small
the identity permutation gives the smallest contribution to the action
in (\ref{itzub}) and therefore is the dominant classical solution.
In this case we use this minimum to evaluate the correlators,
\equation
	I_{IZ}^{-1} \int[dU]e^{N\sum\bar\phi_i\bar\phi_j\vert U_{ij}
	\vert^2}U_{i_1j_1}\ldots
	U_{i_nj_n}U_{k_1l_1}^\ad\ldots U_{k_nl_n}^\ad
	 = \delta_{i_1j_1}\ldots
	\delta_{k_nl_n}S^{i_1\cdots i_n}_{k_1\dots k_n}
	\lab{cor}
\endequation
where we have written the normalized integral over diagonal matrices
\begin{eqnarray}
	&S^{i_1\cdots i_n}_{k_1\dots k_n}=\int\prod_\ell
	 d\theta_\ell \prod_{p<q}
	\sin^2(\theta_p-\theta_q)e^{i\left(\theta_{i_1}+\dots+
	\theta_{i_n}-
	\theta_{k_1}-\dots\theta_{k_n}\right)}/\int\prod_\ell
	d\theta_\ell\prod_{p<q}
	\sin^2(\theta_p-\theta_q) &
\nonumber\\
\nonumber\\
	& =\left\{\matrix{ 1&{\rm~if~} i_1\dots i_n{\rm ~is ~a
	~permutation~
	of~} k_1\dots k_n\cr 0&{\rm~otherwise}\cr}\right.%
\end{eqnarray}
We have decomposed the integration over unitary matrices into an
integration over the diagonals and an integration over the unitary
group modulo diagonals\cite{footn1}. The diagonals are the
`zero modes' for the semiclassical integral and must be integrated
exactly.  The unitary modulo diagonal integral is damped by the
integrand and is performed by substituting the classical
configuration.  Of course, to get the next to leading order the
latter integration must be done in a Gaussian approximation.
It can be done for the first few correlators,
The result is
$$
	 C_{ii}=1-{1\over N} \sum_{k\neq i} {1\over
	{\left(\bphi_i-\bphi_k\right)^2}}
{}~~~{\rm when~}i\neq j~~~C_{ij}={1\over N} {1\over{\left(
	\bphi_i-\bphi_j\right)^2}}
$$

Similar calculations can be easily done for  correlators of more than
two $U$'s
\equation
	C_{ij,kl}=I_{IZ}^{-1} \int [dU] e^{N\sum \bar\phi_i\bar\phi_j
	\vert U_{ij}\vert^2} |U_{ij}|^2 |U_{k \ell}|^2
\endequation
$$
	i=j~~,~~k=\ell~~~~  C_{ii,kk}=1-{1\over N} \sum_{n\neq i} \left[
\frac{1}{(\bphi_i-\bphi_n)^2}+\frac{1}{(\bphi_k-\bphi_n)^2} \right]
$$
\equation
	i\neq j~~,~~k=\ell~~~~~~~~~~~~~~~~~
	~~~~ C_{ij,kk}={1\over N} \frac{1}
	{(\bphi_i-\bphi_j)^2}~~~~~~~~~~~~~~~
\endequation
$$
	i\neq j~~,~~k\neq \ell~~~~  C_{ij,kl}={1\over N^2} \frac{1}
	{(\bphi_i-\bphi_j)^2} \frac{1}{(\bphi_k-\bphi_\ell)^2}
	\left( 1+\de_{ik}\de_{j\ell}
	\right) ~,
$$
where the next corrections will be of the form
$	{1\over N^2}{1\over{\left(\bphi_i-\bphi_j\right)^2}} \sum_{m\neq k}
	{1\over{\left(\bphi_k-\bphi_m\right)^2}}~.
$.  Note also that for the semicircle distribution of the
 eigenvalues one has
$$
	\sum_{j\neq i} {1\over{\left(\bphi_j-\bphi_i\right)^n}}
	= 0~~,~~~~
	{\rm for~~} n>2
$$

\bigskip
\noindent{\bf 2. High Temperature:} This
corresponds to small ${\bar\phi}$. We obtain the correlators by
Taylor expansion:
$$
C_{ij}=\int [dU] \left(1+N\Tr \bar\phi U\bar\phi
U^{\dagger}+\ldots\right)\vert U_{ij}\vert^2=\frac{1}{N}+
\frac{\bphi_i\bphi_j}{N}+\ldots
$$
\vfill
\eject

%\figure{}

\vspace{13cm}

\begin{references}
\bibitem{footn1}Despite the simple form of $S$ its tensor
structure is quite complicated. For example
$$
	S^{i_1i_2}_{k_1k_2}=\delta^{i_1}_{k_1}\delta^{i_2}_{k_2}
	+\delta^{i_1}_{k_2}\delta^{i_2}_{k_1}
	-\delta^{i_1}_{i_2}\delta^{k_1}_{k_2}
$$
\end{references}
\end{document}